\documentclass{emulateapj}

\sfcode`A=1000 \sfcode`B=1000 \sfcode`C=1000 \sfcode`D=1000
\sfcode`E=1000 \sfcode`F=1000 \sfcode`G=1000 \sfcode`H=1000
\sfcode`I=1000 \sfcode`J=1000 \sfcode`K=1000 \sfcode`L=1000
\sfcode`M=1000 \sfcode`N=1000 \sfcode`O=1000 \sfcode`P=1000
\sfcode`Q=1000 \sfcode`R=1000 \sfcode`S=1000 \sfcode`T=1000
\sfcode`U=1000 \sfcode`V=1000 \sfcode`W=1000 \sfcode`X=1000
\sfcode`Y=1000 \sfcode`Z=1000

\slugcomment{{\it To appear in the Astrophysical Journal } }

\shorttitle{Variability of Sgr~A* at 4.5~\micron}
\shortauthors{Hora et~al.}

\usepackage{graphicx}
\usepackage{natbib}
\usepackage{amsmath}

\newcommand{\Sp}{{\it Spitzer\/}}
\newcommand{\SST}{{\it Spitzer Space Telescope\/}}
\newcommand{\Sg}{Sgr~A*}
\newcommand{\ts}{\textsuperscript}

\begin{document}

\title{{\it Spitzer\/}/IRAC Observations of the Variability of \Sg\ and the Object G2 at 4.5~\micron}
\author{
J.~L.\ Hora,\altaffilmark{1}
G.\ Witzel,\altaffilmark{2}
M.~L.~N.\ Ashby,\altaffilmark{1}
E.~E.\ Becklin,\altaffilmark{2,3}
S.\ Carey,\altaffilmark{4}
G.~G.\ Fazio,\altaffilmark{1}
A.\ Ghez,\altaffilmark{2}
J.\ Ingalls,\altaffilmark{4}
L.\ Meyer,\altaffilmark{2}
M.~R.\ Morris,\altaffilmark{2}
H.~A.\ Smith,\altaffilmark{1}
and
S.~P.\ Willner\altaffilmark{1}
}
\altaffiltext{1}{Harvard-Smithsonian Center for Astrophysics, 60
  Garden St., Cambridge, MA 02138 USA}
\altaffiltext{2}{University of California, Los Angeles, CA USA}
\altaffiltext{3}{SOFIA Science Center, Moffett Field, CA USA}
\altaffiltext{4}{Spitzer Science Center, California Institute of Technology, Pasadena, CA 91125 USA}

\begin{abstract}

We present the first detection from the \emph{Spitzer Space Telescope} of 4.5~\micron\ variability from
 \Sg, the emitting source associated with the Milky Way's central
 black hole. The $>$23~hour continuous light curve was obtained with
 the IRAC instrument in 2013 December.  The result
 characterizes the variability of \Sg\ prior to the closest approach
 of the G2 object, a putative infalling gas cloud
 that orbits close to \Sg. The high stellar density at the location
 of \Sg\ produces a background of $\sim$250~mJy at 4.5 \micron\ in each pixel with a large
pixel-to-pixel gradient, but
 the light curve for the highly variable \Sg\ source was successfully
 measured by modeling and removing the variations due to pointing
 wobble.  The observed flux densities range from the noise level of
 $\sim$0.7~mJy rms in a 6.4-s measurement to $\ga$10~mJy. Emission was
 seen above the noise level $\sim$34\% of the time.  The light curve
 characteristics, including the flux density distribution and
 structure function, are consistent with those previously derived at
 shorter infrared wavelengths. We see no evidence in the light curve for activity
 attributable to the G2 interaction at the observing epoch,
 $\sim$100 days before the expected G2 periapsis passage. The IRAC
 light curve is more than a factor of two longer than any previous
 infrared observation, improving constraints on the timescale of the
 break in the power spectral distribution of \Sg\ flux densities.
 The data favor the longer of the two previously published values
 for the timescale.
\end{abstract}

\section{Introduction}

The excellent stability of the \SST\ in its warm mission, combined with recent advances in the modeling of the IRAC instrument response, have opened up new possibilities for studies of \Sg, the emissive source associated with 
the supermassive black hole (SMBH) at the center of the Milky Way. As both the closest example of a SMBH and an extremely under-luminous case, emitting only $10^{-8}$ of its Eddington luminosity, \Sg\ is both a compelling and challenging target to observe. The high precision IRAC photometry obtained for exoplanet studies (\citealt{2012SPIE.8442E..1YI}) suggests that it should be possible to extract a light curve of a faint variable source against a bright, structured background, as is the case for \Sg, which is located in a crowded field of stars and dust at wavelengths accessible to \Sp\ (e.g., \citealt{1996AJ....111.1908S}; \citealt{2004ApJ...601L.159G}; \citealt{2004A&A...417L..15C}; \citealt{2005A&A...433..117V}; \citealt{2006JPhCS..54..176S}; \citealt{2008ApJ...685..958A}). 

High angular resolution ground-based observations offered the first near-infrared (NIR) detections of \Sg, revealing a highly variable source (\citealt{2003Natur.425..934G}; \citealt{2004ApJ...601L.159G}).  However, the duration of continuous measurements, which are essential for accurate modeling of the temporal variability, is limited by the Earth's day-night cycle.  The longest observation reached a maximum  duration of 10 hours only by carefully coordinating observations at the Keck and VLT telescopes (\citealt{2009ApJ...694L..87M}).  \Sp, with its earth-trailing orbit, offers the possibility of much more extended continuous measurement of this enigmatic source, as well as the ability to observe it when ground-based IR telescopes cannot. 

Many studies over the last decade have focused on characterizing the temporal properties of \Sg\ at NIR wavelengths.  
Early studies with limited time baselines suggested the possibility of a 20-minute quasi-periodic variation, which held the hope of revealing the spin of the black hole (e.g., \citealt{2003Natur.425..934G}).  Subsequent analyses of the statistical properties of more extensive data,  however, have shown that the NIR variability of \Sg\ is well represented as a continuous, red-noise process in which the power
spectral density (PSD) follows an inverse power-law dependence on temporal frequency
(\citealt{2009ApJ...691.1021D}).  The PSD power law extends from
frequencies corresponding to tens of seconds down to a break
frequency corresponding to hundreds of minutes
(\citealt{2009ApJ...694L..87M}, \citealt{2012ApJS..203...18W},
\citealt{2014arXiv1403.5289M}), a timescale difficult to sample from the ground owing to the limited duration of uninterrupted light curves, which typically extend for only $\sim 3$ to 6 hours.  

While the physical processes underlying the variability of the emission have yet to be identified, many candidate processes have been suggested. These include magnetic reconnection events or shocks in an inhomogeneous accretion flow, adiabatically expanding plasma blobs, intermittent jets or unstable jet shocks, and multiple orbiting and evolving hot spots  (\citealt{2012A&A...537A..52E}, \citealt{Yuan2011}, \citealt{2008A&A...479..625E}, \citealt{2008JPhCS.131a2008Z}). 
 Although each of these models explain individual aspects of the NIR variability process, there is no model that matches all the properties of the variability as we observe it. Those properties are:

\begin{itemize}

\item The process is random. Any valid model should explain not only single instances of outbursts, but needs to be statistical in  nature \citep{2009ApJ...691.1021D, 2009ApJ...694L..87M, 2012ApJS..203...18W}.
\item The process has one state \citep{2014arXiv1403.5289M, 2012ApJS..203...18W}.
\item The process is continuous in time \citep{2009ApJ...691.1021D, 2014arXiv1403.5289M, 2012ApJS..203...18W}.
\item The process has no characteristic flux density within the observable flux density range. In particular, there is no evidence for a quiescent state in the NIR (\citealt{2012ApJS..203...18W}).
\item The NIR emission is polarized \citep{2006A&A...455....1E}.
\item The process has a constant NIR spectral index that does not vary with flux but does slightly vary with time (\citealt{2007ApJ...667..900H}; Witzel in prep.).
\item The process has a characteristic timescale of several hours \citep{2009ApJ...694L..87M, 2012ApJS..203...18W}.
\item The NIR flux density potentially correlates with the variability in the X-rays and in the sub-mm regime \citep{2004A&A...427....1E, 2008A&A...492..337E, 2013arXiv1308.5968D}.

\end{itemize}

 The only full general relativistic magnetohydrodynamic (GRMHD) simulation of the accretion disc around \Sg\ that satisfies several of the above criteria (in particular the statistical nature and a constant spectral index that matches the observed value) and that is able to reproduce the observed flux range is the one by \citet{2013MNRAS.432.2252D}. Their resulting light curve \citep[Figure 15 in][panel 2]{2013MNRAS.432.2252D} exhibits a timescale short enough to produce several 20 mJy outbursts a day. In order to determine if this (or any) model accurately describes the processes around \Sg,  accurate measurements of the timescales are required.  Comparisons of break timescales across the electromagnetic spectrum, such as those measured at radio wavelengths by \cite{2013arXiv1308.5968D}, offer important additional constraints on models for the physical processes underlying the variability of the emission.

Observations with \Sp\ can greatly extend the duration of continuous light curves, which is key for measuring the NIR break timescale with sufficient accuracy to compare to the timescales measured at other wavelengths. Well-characterized light curves for \Sg\ are also particularly important at this time due to the recent report of a putative $3\;\rm{M}_{\earth}$ gas cloud -- G2 -- that appears to be plunging toward the SMBH at the Galactic center with a predicted closest approach of less than 3000 times the event horizon \citep{2013ApJ...774...44G,2013arXiv1312.1715M}.  If G2 is indeed a gas cloud, it will be ripped apart by the tidal forces of the SMBH during closest approach and then mostly accreted (\citealt{2012ApJ...750...58B}; \citealt{2012ApJ...755..155S}; \citealt{2012ApJ...759..132A}; \citealt{2013arXiv1311.4507S}; \citealt{2014MNRAS.440.1125A}).   While the identification of the source is controversial, and many alternative models containing a central stellar source have been proposed (e.g., \citealt{2012NatCo...3E1049M}; \citealt{2012ApJ...756...86M}; \citealt{2013A&A...551A..18E}; \citealt{2013ApJ...768..108S}; \citealt{2013ApJ...776...13B}), observations indicate that low density gas associated with G2 is being tidally disrupted (\citealt{2012Natur.481...51G, 2013ApJ...763...78G, 2013ApJ...774...44G}; \citealt{2013ApJ...773L..13P}). Most models predict that the disrupted material should eventually be accreted onto the SMBH (see also \citealt{2014arXiv1401.0553F}) although the amount and timing of the increased accretion rate are highly uncertain. \Sp\ observations offer a window into \Sg's short timescale variability that is comparable to that obtainable from high-angular-resolution ground-based observations.

Motivated by the need for longer light curves of \Sg\ in order to
define the NIR PSD break and to determine the effects on its emission from the accretion of G2, 
we have undertaken a study with the Infrared Array Camera
(IRAC; \citealt{2004ApJS..154...10F}) on the \SST\ (\citealt{2004ApJS..154....1W}).  This paper reports our first 23.4-hour light
curve of \Sg\ at a wavelength of 4.5~$\mu$m and demonstrates the
capability of this instrument for monitoring variable emission of a
 source in a crowded region. Our present and upcoming
observations (see \S\ref{s:discussion}) are timed to bracket the close periapsis passage of the G2 object as it orbits close to \Sg.  These
observations should delimit any excess emission that might be
produced by enhanced accretion or G2 gas interaction with the
existing accretion flow. The observation described here took place
$\sim$100 days prior to the G2 periapsis passage, which was expected to occur in
2014 March (\citealt{2013arXiv1312.1715M}; \citealt{2013ApJ...773L..13P};
\citealt{2013ApJ...774...44G}).

\section{Observations and Data Reduction}
\label{obs_section}
\begin{figure*}
\begin{center}
\includegraphics[scale=0.69, angle=0]{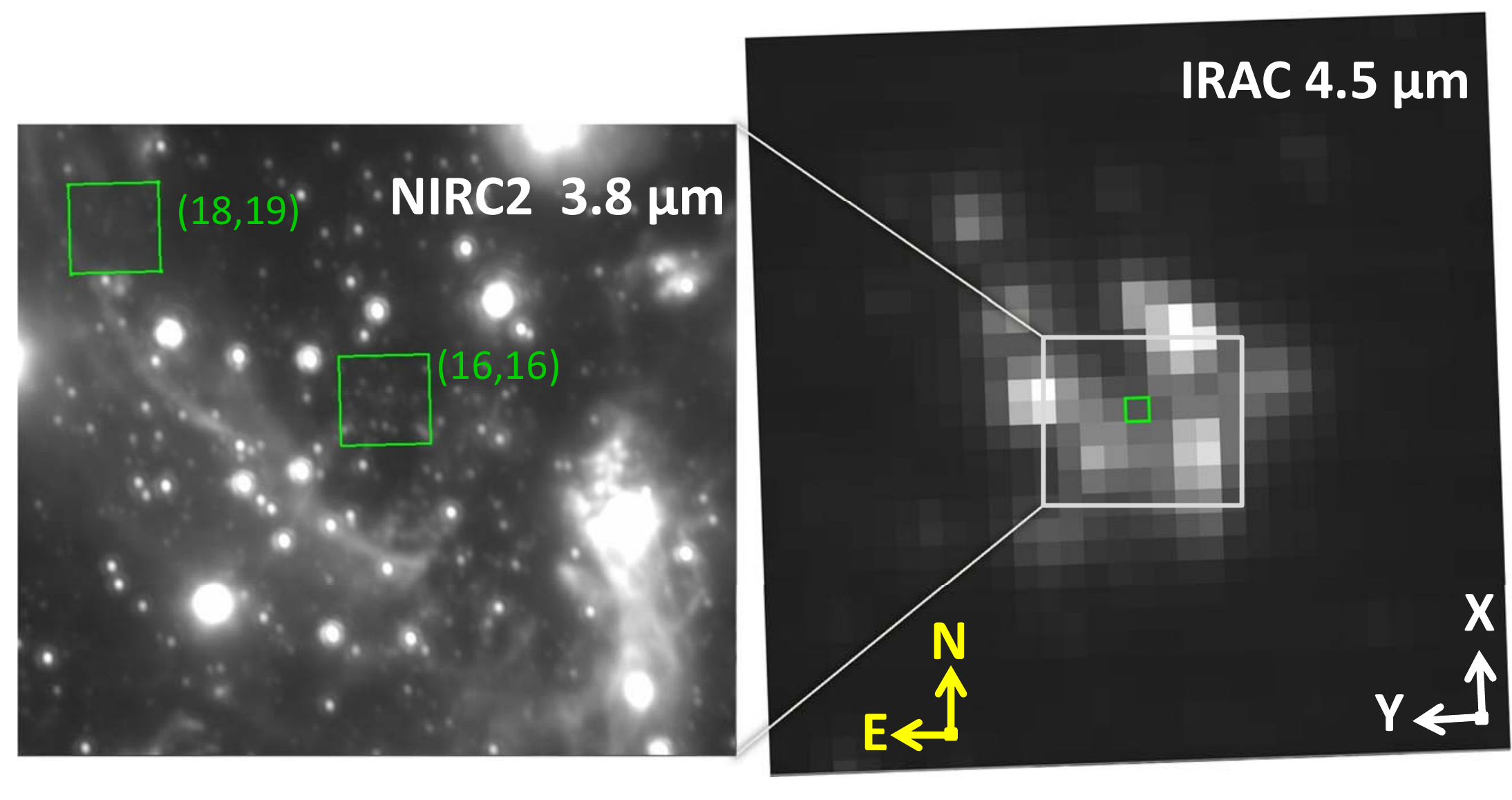}
\caption{ The field near \Sg\ in the infrared. North is up and
  East to the left in both images.  Left panel: A diffraction-limited image at  3.8~$\mu$m obtained at the Keck observatory in 2013 April with
  NIRC2. The region shown is $\sim$10\arcsec$\times$9\arcsec\ in size. The green box near the center indicates the location and
  size of  IRAC pixel (16,16). \Sg\ itself is located near the center of the
  box. The green box in the upper left indicates the reference pixel
  (18,19).  Right panel: The 39\arcsec~square  IRAC 4.5~$\mu$m mosaic used as the reference image to determine the 
relative frame set pointing  offsets.
  The green box indicates pixel (16,16), 
  and the light gray box
  indicates the approximate field of the Keck image in the left panel.  The directions of
  positive X and Y on the IRAC subarray are indicated by arrows; the
  position angle of the IRAC observations was $\sim$91\fdg7 east of
  north. }\label{f:map}
\end{center}
\end{figure*}

Observing the variability of \Sg\ with IRAC presents a challenge
because the source is surrounded by several much brighter infrared sources
and extended emission (\citealt{1989AJ.....98..204T, 1995ApJ...439..682C,
1997MNRAS.291..219G, 2002ApJ...577L...9H}; see Figure~\ref{f:map}). The central source
complex is unresolved at IRAC's angular resolution (1\farcs43~FWHM at
4.5~$\mu$m for stars centered on a pixel --- \citealt{2004ApJS..154...10F}), and
the surface brightness is a strong function of position.
Therefore any change in the telescope pointing, even at the sub-pixel
level, produces a relatively large change in the signal on a
given pixel.  \Sp's superb pointing stability minimized this effect; however, 
correcting for the remaining pointing-induced variations was the
main task of the data analysis, and the need for this correction was
taken into account in planning the observations.

All observations in this \SST\ program (ID\#10060) used IRAC in
subarray mode, which obtains 64 consecutive images (a ``frame set'')
of a 32$\times$32 pixel region (1\farcs21 per pixel)  near the corner
of the 256$\times$256 pixel array. The observations were conducted as
three custom ``Instrument Engineering Requests'' (IERs) because the
standard IRAC Astronomical Observation Template does not allow the
observation sequence we designed for this program.  We used the ``PCRS Peakup'' mode to 
position \Sg\ as close as possible to the center of pixel (16,16).
The detailed design of the IERs is described in Appendix A.

For our data reduction we used the standard Basic Calibrated Data
(BCD) products (version S19.1.0) downloaded from the \Sp\ Heritage
Archive\footnote{The Spitzer Heritage Archive
  (http://sha.ipac.caltech.edu/) is part of the NASA/IPAC Infrared
  Science Archive operated by the Jet Propulsion Laboratory,
  California Institute of Technology under a contract with the
  National Aeronautics and Space Administration.}. Because we expect
no detectable variation from \Sg\ on a time scale of  6.4~s (based on extrapolations of the source
characteristics observed in the $K$-band and the noise level of the IRAC observations
at 4.5~\micron), the first step in the
reduction was to combine each frame set into a single 32$\times$32
pixel image or ``6.4-s BCD coadd''. This was done using the {\tt
  subcoadd\_bcd} routine in the IRACproc image processing software
(\citealt{2006SPIE.6270E..65S}), which performs an average of the frames with
outlier rejection to eliminate cosmic rays.

\begin{figure*}
\begin{center}
\includegraphics[scale=0.6, angle=0]{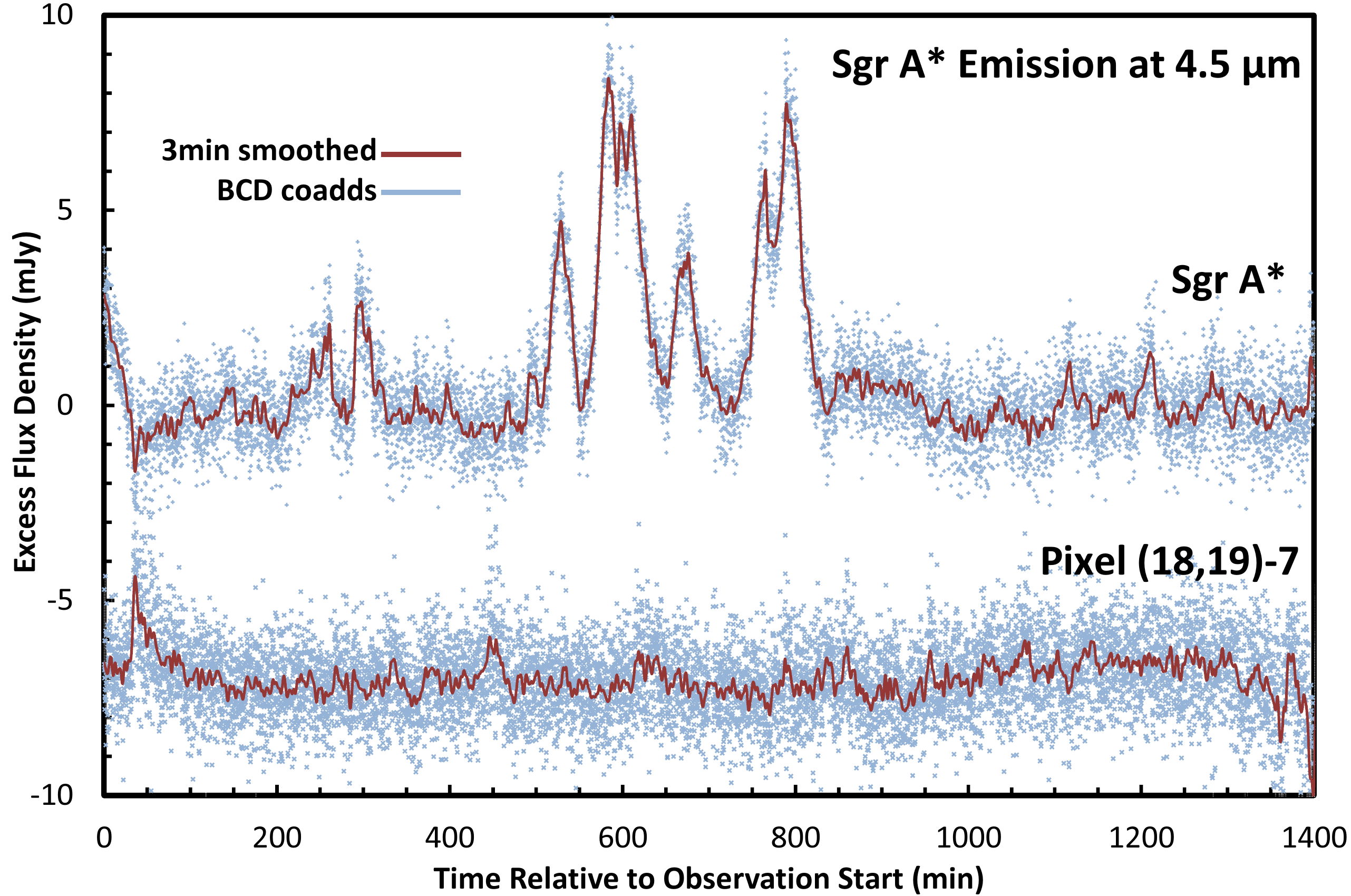}
\end{center}
\caption{Excess flux density for \Sg\ and for the reference pixel in
  mJy.  Blue points show the flux density for each 6.4-s BCD coadd,
  and the red lines show a running average smoothed with a Gaussian
  kernel having a 3~minute FWHM. The upper line and blue crosses are
  for \Sg, and the lower line and blue Xs are for reference pixel (18,19)
  offset vertically by $-7$~mJy.  The calibrated values plotted are
  the difference between the observed value of the pixel in the 6.4-s
  BCD coadds and the predicted value for the
  measured $(X,Y)$ offset of each coadd , as described in Appendix A. Pixel
  values have been corrected to total flux density by the
  position-dependent ratio of total flux density to central-pixel
  signal.  The horizontal axis shows the time in minutes relative to the start
  time of the first monitoring 6.4s BCD coadd. }\label{sgrAplot}
\end{figure*}

We extracted the flux density of \Sg\ for each BCD coadd as described in Appendix A.
The resulting light curve is shown in the upper part of
Figure~\ref{sgrAplot}.  
Because we have fit the signal in the
relatively quiescent periods during the monitoring, we cannot derive
the baseline flux from \Sg\ itself, only the excess above the level
that persisted for a period shorter than the 23.4 hours that we
monitored the source. Several significant peaks above the baseline
level are seen, the largest $\sim$10~mJy.  The rms variation of the
6.4-s BCD coadds for the time period 850--1400 minutes, which is
relatively free of large peaks, is 0.9~mJy.  There is also a
deviation from the model fit in the first 100 minutes that appears
different from the subsequent variations.  This might indicate some
systematic error in our fit at the beginning of the time series.  We
will be able to investigate this possible effect further after we
obtain more monitoring observations scheduled for 2014 June--July.

\begin{table}
\begin{center}

\caption{Sgr A* Light Curve Data}
\begin{tabular}{ccc}
\tableline
\tableline
&&\\
Observation  & Sgr A*& Reference\\
Date\tablenotemark{a} & Flux Density&Flux Density\\
(MJD)&(Jy)&(Jy)\\
\tableline
&&\\
56636.1808736&  0.00210  &0.00036 \\
56636.1809705 & 0.00404 & 0.00228 \\
56636.1810678  &0.00393 & 0.00001 \\
56636.1811650 & 0.00294 & 0.00021 \\
56636.1812623 & 0.00233 & 0.00096 \\
56636.1813596 & 0.00299 &-0.00080 \\
56636.1814566 & 0.00338 &-0.00046 \\
56636.1815536 & 0.00215 & 0.00025 \\
56636.1816511 & 0.00276 &-0.00028 \\
56636.1817483 & 0.00239 &-0.00058 \\
\tableline
\end{tabular}
\tablenotetext{1}{ Modified Julian Date (solar system barycenter) of the midpoint of the 6.4s BCD coadd.\\}
\tablecomments{The flux density values in this table are plotted in Figure~\ref{sgrAplot}.  See Appendix A for a description of how they were derived from the observations.\\
(Table~\ref{LightcurveData}  is available in its entirety in a machine-readable form in the online journal. A portion is shown here for guidance regarding its form and content.)}\label{LightcurveData}

\end{center}
\end{table}

 As a test of our reduction method, we extracted the output of 
a reference pixel (18,19) in the same way as for (16,16), also shown in Figure~\ref{sgrAplot}. 
There is one peak in the extracted flux density
for the reference pixel near the beginning of the time series that rises to
$\sim$2.5 mJy in the smoothed data, but other fluctuations are
smaller, and the rms variation is similar to that for \Sg\ pixel (16,16).  The light curve data for \Sg\ and the reference
pixel (18,19) are given in Table~\ref{LightcurveData}.

\section{Variability properties of \Sg\ at 4.5~$\mu$m}\label{timing}

The statistics of the \Sg\ NIR variability have been well
characterized on the basis of a decade of 2.1--2.2~$\mu$m ($K$-band)
observations with AO instruments at the VLT
\citep{2011ApJ...728...37D,2012ApJS..203...18W} and Keck
\citep{2009ApJ...691.1021D} observatories and with the two observatories combined
\citep{2009ApJ...694L..87M}. The main properties are:

\begin{itemize}

\item The \Sg\ spectral index $\alpha$ ($f_\nu \propto
  \nu^{-\alpha}$) measured between 1.6 and 3.7~$\mu$m is $0.6 \pm
  0.2$, where the quoted uncertainty includes the uncertainty in the
  differential extinction and the possibility of small variations in
  $\alpha$ as $f_{\nu}$ varies. The spectral index variations have been observed to be uncorrelated
with the flux density of the source, and they are small enough to be disregarded in our analysis (\citealt{2007ApJ...667..900H} and references
  therein, Witzel et al. in prep.).

\item The probability distribution of the intrinsic (reddening corrected) flux density $f_{K}$ at
  2.2~$\mu$m is skewed, i.e. it shows a low median value ($\sim$1 mJy) and a  tail toward high flux densities
(flux densities as high as 30 mJy have been measured). The flux density distribution (FDD)
can be described by a power law with index
  $\beta = 4.22$ and a normalization $f_{0K} = 3.57$~mJy \citep[][eq.~9]{2012ApJS..203...18W}:
  \begin{equation}
\begin{split}
  P(f_K) = [(\beta-1)/(f_{0K})][(f_K+f_{0K})/(f_{0K})]^{-\beta},\qquad f_K\ge0\\
  P(f_K) = 0,\qquad f_K<0 .
\end{split}
  \label{eq:fd}
  \end{equation}
  The cutoff at zero flux density makes the power-law
  normalizable and guarantees that the intrinsic flux
  density of the source is never negative.
% The cutoff flux density $f_{\rm min}$ represents the measurement
%  noise and is ideally $f_0$ but can be different if the measurement
%  noise is non-Gaussian.
%  The breakpoint of this power-law
%  (required to make the power-law normalizable) is at
%  $f^{*}_{min} = 0$~mJy, guaranteeing all possible values for the
%  flux density to be positive (\citealt{2011ApJ...728...37D}.

\item The NIR variability shows a so-called rms--flux relation, i.e., the
  typical amplitude of variation is proportional to the flux density at
  which the variation occurs (\citealt{2012ApJS..203...18W}).  
The variability is continuous and shows no
  indication of a quiescent state or different variability states
  (\citealt{2012ApJS..203...18W}, \citealt{2014arXiv1403.5289M}).  

\item The PSD of the variability is a
  red-noise power-law spectrum with
  break at a temporal frequency $f_{bK}$ corresponding to a few
  hundred minutes:
  \begin{equation}\label{psddef}
  \begin{split}
  {\rm PSD}(f_t) &\propto f_t^{-\gamma_1},\qquad f_t>f_{bK}\\
  {\rm PSD}(f_t) &\propto f_{bK}^{\gamma_0-\gamma_1}f_t^{-\gamma_0},\qquad f_t\le f_{bK}
  \end{split}
  \end{equation}
The value of $f_{bK}$ corresponds to
 $154^{+124}_{-87}$~minutes according to  \cite{2009ApJ...694L..87M} or
 to  100--1000~minutes according to  \cite{2012ApJS..203...18W}.
The power-law index for long time intervals is $\gamma_0 \approx 0$ and for short time intervals is $\gamma_1 \approx
2.0$ . \cite{2012ApJS..203...18W} also tested PSD models with a second break according to equation (\ref{psddef}) and the additional equation:
\begin{equation}\label{psddef2}
{\rm PSD}(f_t) \propto f_{bK,2}^{\gamma_2-\gamma_1}f_t^{-\gamma_2},\qquad f_t\le f_{bK,2}\; \; ,
\end{equation}
and  $f_{bK,2}=0.05$, which resulted in higher acceptance values but not in a significant improvement that justified the additional parameters. 

\end{itemize}

The goal of our statistical analysis is to investigate whether the
4.5~$\mu$m light curve can be understood with
the existing models for the FDD and the PSD. We applied methods similar to
those of Witzel et al.\ but modified them to account for: 

\begin{itemize}

\item higher
  measurement noise and different baseline flux density level owing to
  unresolved  nearby sources and the data reduction method

\item having a single 23h sample during which the light curve maintains
  significant correlation from beginning to end.  (The ground-based
  observation periods were shorter individually, and were separated by months
  or years and therefore have no correlation from one monitoring
  campaign to another.)

% \item a constant scaling factor to represent the spectral index
%   from 2.2 to 4.5~$\mu$m, the differential extinction between the two
%   wavelengths, and potential flux density calibration errors.

\end{itemize}

We analyzed the data in two steps, first looking at the FDD and
second at the timing properties of the \Sp\ data. Both steps 
made use of equation (28) of \citet{2012ApJS..203...18W}, which allows us
to transform a unit normal-distributed random variable $y$ into a
power-law distributed random variable $T(y)$ that takes on values $0 <
T(y) < \infty$:
\begin{equation}
T(y) =  f_{\rm{0}} \cdot \left\lbrace \frac{1}{2}\left[ 1 +
    {\rm{erf}}\left(\frac{y}{\sqrt{2}}\right)\right]\right\rbrace^{\frac{1}{\left(1-\beta
    \right)}} - f_{\rm{0}}~~. 
\label{transf}
\end{equation}
where erf is  the Gaussian error function,  the power-law index
$\beta = 4.22$, and $f_{0}$ is the normalization flux density at the wavelength observed.

\subsection{Flux density distribution} \label{cdfsec}

Even an observation duration of $\sim$23~h may not be vastly longer
than the coherence timescale, and its FDD is only a single
sample of the distribution of the variability process. As a
consequence, the estimate of the FDD power-law index $\beta$ is
uncertain even if the data are known to be drawn from a power-law
distribution. Therefore, instead of deriving the power-law parameters
independently from our dataset, we have adopted the parameters
measured in the $K$-band \citep{2012ApJS..203...18W} and determined whether the new dataset is a
likely realization of the same random process.
The metric of comparison is the Kolmogorov-Smirnov ($KS$) statistic,
for which we derived acceptance levels after establishing the
timing properties by Monte Carlo simulations as described in
Section~\ref{s:timing}. 

We determined the $KS$ value in the following way: 

\begin{itemize}

\item Due to the measurement techniques described above in \S\ref{obs_section}, the empirically measured 4.5~\micron\ flux densities $f$ omit a component
  corresponding to the average flux density when \Sg\ is in its relatively
  quiescent periods. This extra component is represented by a constant
  $c$, and we 
% For a given background flux $B$, which includes the average
%   quiescent flux of \Sg\ plus unresolved nearby sources, we
calculated the empirical complementary (i.e., $P(f>x)$ rather than
$P(f<x)$) cumulative distribution function ${\rm CDF}(f + c)$ from
the observed light curve.

\item For comparison with the measured light curve, we generated
  $10^{7}$ unit normal-distributed random numbers and 
  transformed them into power-law distributed numbers according to
  equation~(\ref{transf}). The resulting values were then multiplied by a
  factor $s$ (to scale the dereddened $K$-band flux density values to the observed 4.5~\micron\ flux density)
to represent simulated values of $(f+c)$ at 4.5~\micron.
  To these we added Gaussian white noise with a standard deviation 
  $\sigma = 0.7$~mJy. This value, rather than the empirical standard
  deviation  $\sigma = 0.9$~mJy, is justified by the low-time-lag value of the
  structure function derived in the next section 
  and is consistent with intrinsic  fluctuations of \Sg\ of
  $\sim$0.6~mJy rms at the noise dominated flux levels.  The result is simulated
values of $f$ with the constant $c$ added, i.e, ($f + c$).

\item From the simulated light curve, we calculated the
  complementary cumulative 
  distribution function ${\rm{CDF}_{\rm{sim}}}({f+c})$ and resampled based on a linear
interpolation to ensure the CDF  flux densities are the same for all CDFs. Then
we calculated
  \begin{equation}
    KS = \max_{f > f_{\rm{min}}}\left[CDF(f+c)-CDF_{\rm{sim}}({f+c})\right]
    \label{KS}
  \end{equation}
  where the parameter $f_{\rm{min}}$ separates the part of the CDF that is
  noise dominated from the part   actually represented by a power law.
\end{itemize}

\begin{figure}[ht]
   \centering
   \includegraphics[width=8.75cm]{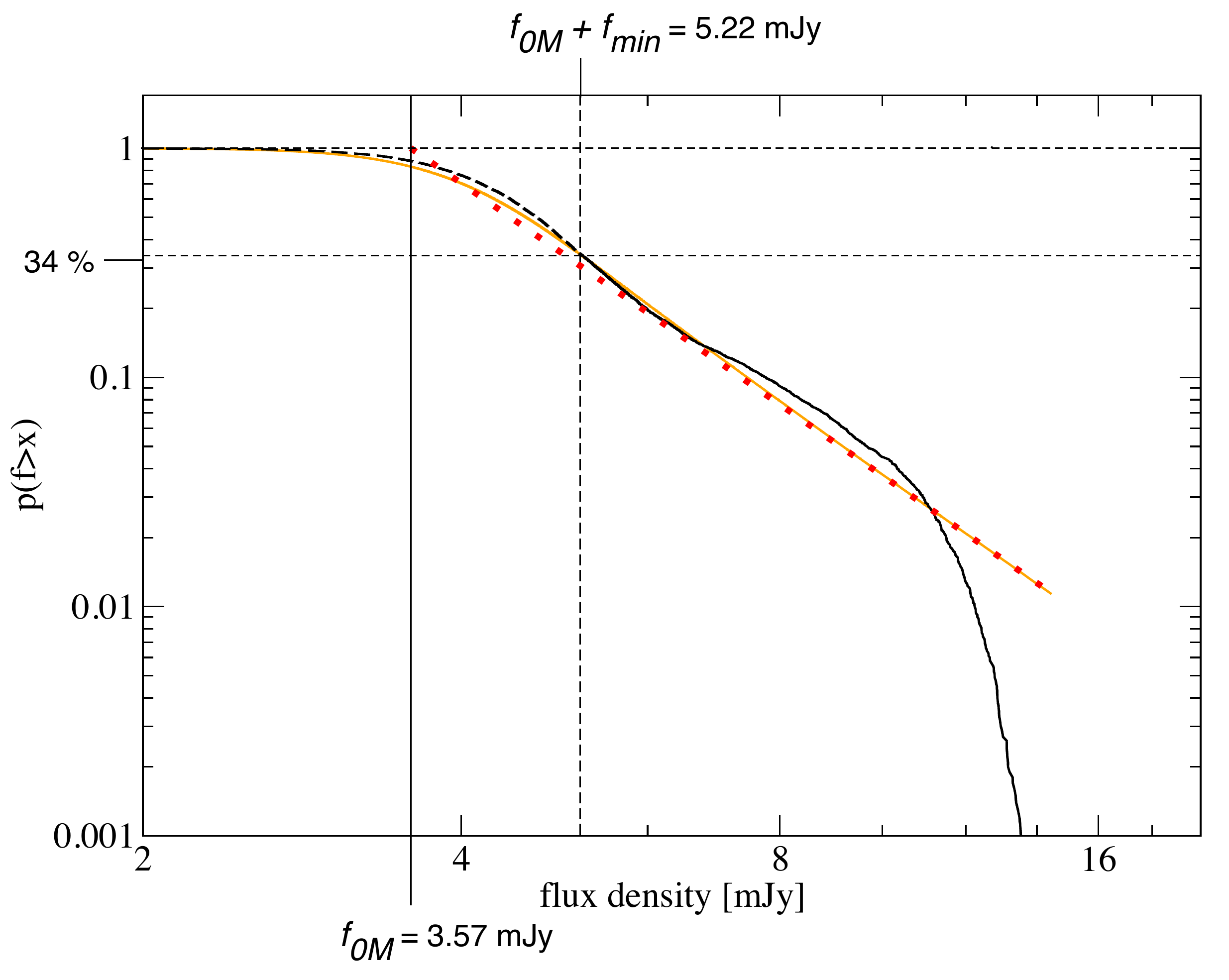}
   \caption[Complementary cumulative distribution function.]{Complementary cumulative
     distribution function and power-law fit to the data. The CDF
     of the light curve data $f$ is shown in black. The solid part of
     the line represents flux densities that were used for the
     $KS$-value estimation, and the dashed black line is the part
     dominated by measurement noise. The simulated CDF for the best
     parameters is in orange, and the red squares show the power-law
     model adopted from pre-existing $K$-band data
     \citep{2012ApJS..203...18W}. The plot abscissa is $f + f_{0M}$ (with $f_{0M} = s \cdot f_{0K}$) in
     order to show the noise-dominated region of the data and to make
     the power law display as a straight line.  The vertical lines show the power law normalization $f_{0M}$
     (3.57~mJy) and the flux density ($f_{0M}+f_{min}=5.22 \mu$Jy) at which the simulated CDF (orange) begins to deviate
     from the power law (red) due to measurement noise.}
         \label{cdf}
\end{figure}

The final calculation consisted of minimizing the $KS$ value over the
parameters $c$, $s$, and $f_{\rm{min}}$.\footnote{
If the distribution of the
measurement noise were exactly known, it would not be necessary to
restrict the CDF comparison  to
values  ${>}f_{\rm{min}}$. But because the noise is caused
by residuals in the subtraction of nearby sources, it is most likely
not Gaussian. Including flux densities with $f <
f_{\rm{min}}$ would lead to the $KS$ estimate being dominated by the
insufficiently known measurement noise. Nevertheless it is
important to include measurement noise in the model because noise
creates  deviations from a 
power law even for flux densities with $f > f_{\rm{min}}$ (as
displayed in Figure~\ref{cdf}). }
The result is shown in Figure~\ref{cdf}. The best parameters are
$c = 0.94$~mJy, $s=1.0$, and $f_{\rm{min}} = 1.65$~mJy, and the
resulting best $KS$ value is $0.0133$ corresponding to a 1.3\%
maximum difference between the real CDF and the one generated from
the best-fit model.  The corresponding probability cannot be looked
up in standard tables because the data are correlated but can be
derived from simulations as discussed in Section~\ref{s:timing}.

\begin{table*}
\begin{center}
\caption{PSD models and their probabilities}
\begin{tabular}{l|cccc}
\tableline
\tableline
&&&&\\
& PSD 1 & PSD 2 & PSD 3 & PSD 4 \\
&&&&\\
\tableline
&&&&\\
$f_{bK}$ & 0.0065 $\rm{min}^{-1}$&0.0065 $\rm{min}^{-1}$ & 0.0018 $\rm{min}^{-1}$ & 0.0018 $\rm{min}^{-1}$ \\
$f_{bK,2}$&\nodata&0.05 $\rm{min}^{-1}$&\nodata&0.05 $\rm{min}^{-1}$\\
$\gamma_{1}$&2.0&2.0&2.0&2.0\\
$\gamma_{2}$&\nodata&3.8&\nodata&3.8\\
$p(\rm{CDF})$&0.1110&0.2032&0.1697&0.1750\\
$p(\chi^{2}\vert\rm{CDF}) $&0.0032&0.0261&0.0488&0.1033\\
$p(\chi^{2}\vert\rm{CDF}) \cdot p(\rm{CDF})$&0.00036&0.00530&0.00828&0.01807\\
freq. &1/2814&1/189&1/121&1/55\\
$p(\chi^{2}_{\rm{ps}})$&0.41&0.44&0.74&0.73\\
\\
\tableline
\end{tabular}
\tablecomments{Parameters and derived probabilities of the four PSD models discussed in the text. The break timescales and slopes are as defined in Eqs.~\ref{psddef} and \ref{psddef2}. The parameter $p(\rm{CDF})$ is the fraction of light curves that match the CDF of the observed data, as described in detail in Appendix B. Within these, $p(\chi^{2}\vert\rm{CDF})$ is the fraction of light curves that have a larger $\chi^{2}$-values than the observed data. The likelihood function of the parameter set is the product of both probabilities. freq. is the corresponding frequency of occurrence. The $p(\chi^{2}_{\rm{ps}})$-value is the fractions of structure functions with larger $\chi^{2}$-values than the observed data, derived from light curves without CDF constraints.}\label{PSDparams}
\end{center}
\end{table*}

\begin{figure}[ht]
   \centering
   \includegraphics[width=8.7cm]{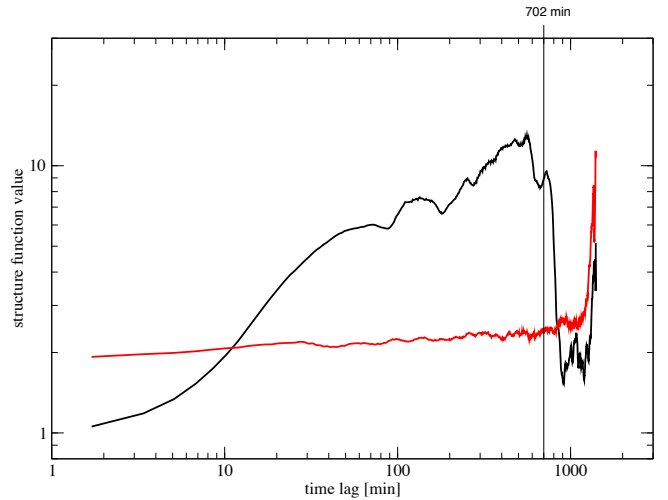}
      \caption[Structure function.]{Structure function of the light
        curve of \Sg\ (black) and of the comparison pixel (red). The
        vertical line at 702~minutes marks half of the observing
        window duration.  See \S\ref{s:timing} for more details.} 
         \label{struct}
\end{figure}

\begin{figure*}[ht]
   \centering
   \includegraphics[width=14cm]{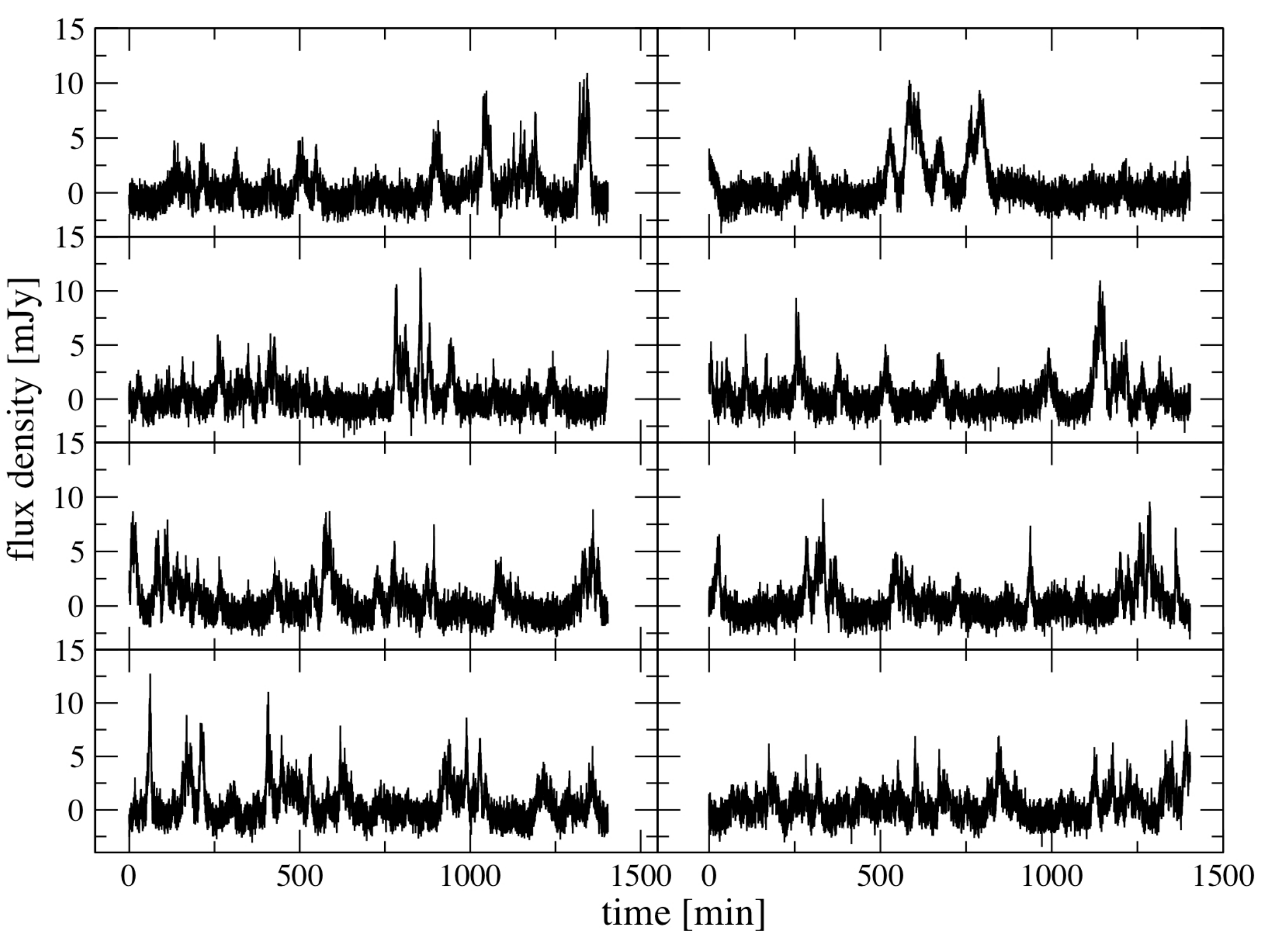}
   \caption{Real and simulated light curves. The upper
     right panel shows the observed data, and other panels show
     simulated light curves based on the best FDD model and
      a PSD-model with a break timescale at 154~minutes (model 1 in Table \ref{PSDparams}). }
         \label{lc}
\end{figure*}

\subsection{Timing analysis}\label{s:timing}

Timing analysis is about understanding the statistics of flux density
differences between measurement points separated by a given time
interval.  A natural metric for characterizing
timing properties is  the first-order
structure function, which measures the
mean value of the flux density variance for a given time lag
$\tau $.  It is defined as (\citealt{1985ApJ...296...46S,
  2009ApJ...691.1021D}):
\begin{equation}
V(\tau) = \langle\left[f\left(t + \tau \right) - f(t) \right]^{2} \rangle~~.
\end{equation}
The structure function is the suitable metric in the case of data with unequal sampling, large gaps, or an observation window not vastly larger than the coherence timescale, which are sampling properties that introduce biases to standard Fourier techniques. The ultimate goal is to determine the shape of the PSD of the underlying process. Computing the PSD from a given structure function is not a trivial task. Only for very idealized cases can an analytical expression be used \citep{1985ApJ...296...46S,2010MNRAS.404..931E}. In general, Monte Carlo simulations that use the appropriate window function have to be used. The method applied here derives structure functions for a given PSD in a Monte Carlo simulation and determines the probability of the observed structure function and is very similar to the methods described by \cite{2002MNRAS.332..231U}, \cite{2009ApJ...691.1021D}, and \cite{2009ApJ...694L..87M}. 
The observed structure function of our 4.5~\micron\ light curve is shown in
Figure~\ref{struct} with a time-lag binning of 1.7 minutes\footnote{As explained later in the discussion, the \Sp\ data rebinned to about 0.85 minutes have approximately the same $S/N$ as the VLT/NACO light curves. A time lag of 1.7 minutes corresponds to the Nyquist frequency at this $S/N$ level.}. While the values for \Sg\ cover more than one
order of magnitude, the structure function of the comparison pixel is
almost constant, showing that the measurement noise is close to
white.

The observed light curve (Figure~\ref{sgrAplot}) has its brightest
peaks occurring close to the middle of the light curve with long,
almost featureless stretches on both sides.  This is reflected in the
structure function by a maximum near 700~minutes followed by a steep
decline at longer time lags.  However, time lags larger than half the  $\sim$1400~minute light curve
duration have large uncertainties because progressively fewer flux density points contribute to the structure
function.  (For shorter time lags, all the flux density points in the light curve
contribute to every structure function value though in different
combinations.  The limit is that only two light curve points
contribute to  the structure function  at the full light curve duration.)

\begin{figure*}[ht]
   \centering
   \includegraphics[width=14cm]{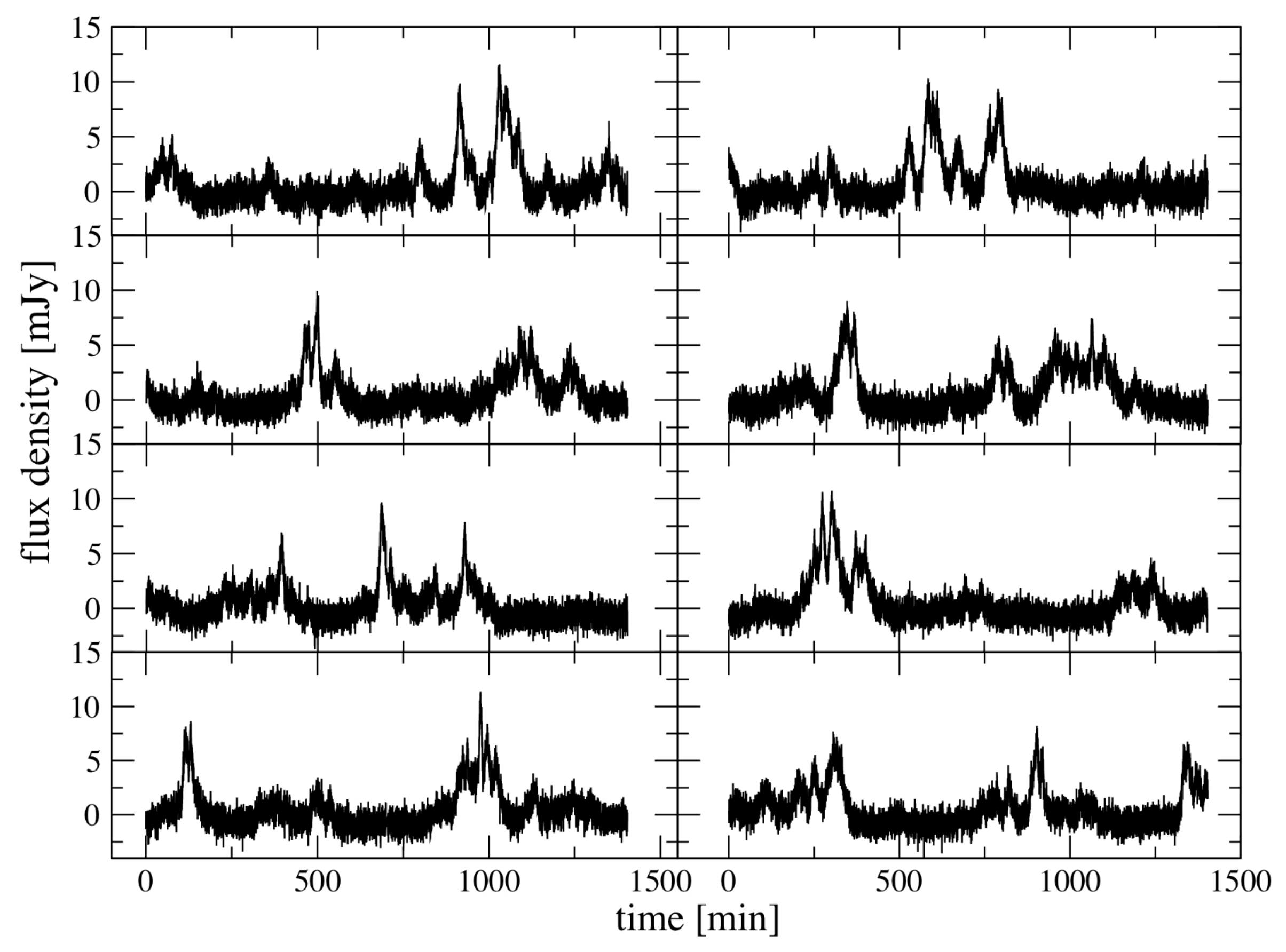}
   \caption{Real and simulated light curves.  The
     upper right panel shows the observed data, and other panels show
     simulated light curves based on the best FDD-model and
     a PSD model with a break timescale at 560~minutes (model 4 in Table \ref{PSDparams}).}
         \label{lc2}
\end{figure*}

The formal way to infer best-fit parameters of the PSD (including uncertainties) from our data would be to use the likelihood function $p(\chi^{2}\vert\rm{CDF}) $ to derive the PSD parameters (slopes, break timescales, and their errors) by maximizing the likelihood function. Due to the high computational demands of the simulation process described in Appendix B and with only one 23.4h dataset, our goal here is to illustrate the advantages of such long, continuous monitoring rather than to explore the full parameter space. We therefore developed  a pseudo $\chi^{2}$ ($\chi^{2}_{\rm{ps}}$) metric to quantify the difference between structure functions, and the likelihood functions  $p(\chi^{2})$ and $p(\chi^{2}\vert\rm{CDF})$. The result quantifies the relative probability of individual PSD models, as described in detail in Appendix B.  We restricted our analysis to four selected PSD models (listed in Table \ref{PSDparams}), which represent the following scenarios discussed in the literature: 1) a timescale of 154 minutes \citep{2009ApJ...694L..87M, 2014MNRAS.442.2797D}; 2) same as model 1 but with an additional timescale in the range of the debated quasi-periodic oscillation \citep[20 minutes;][]{2003Natur.425..934G}; 3) a timescale of 556 minutes, significantly larger than expected for orbital timescales but comparable to the timescale found in the sub-mm \citep{2014MNRAS.442.2797D} and indicated by the analysis of \citet{2012ApJS..203...18W}; and 4) same as model 3 but with the 20 minute timescale added. 

The results of our simulations are given in Table~\ref{PSDparams}.
For the single-break PSD model  1,  $10\%$ of the simulated light
curves have an FDD in accordance with our constraints, and $0.4\%$ of
these accepted light curves show modified $\chi^{2}$ values of the
magnitude of the observed structure function. This means that one in about
2800 light curves has  the observed
FDD and structure function (or one 23h stretch every 7.8 years). In addition, this
model does not explain the maximum of the structure
function at a time lag of about 560~minutes. In contrast,
the double-break PSD model  4 (with breaks at 20 and 560 minutes) produces $18\%$ of the simulated light
curves having an FDD in accordance with our constraints. Fully $10\%$ of these
show  $\chi^{2}_{\rm{ps}}$ values of the
magnitude of the observed structure function. This means that about
one in 55 light curves has the observed
FDD and structure function (or one 23h stretch every 8 weeks). Model 4 also produces confidence
intervals that enclose the observed structure function for all time lags.

PSD models 2 and 3 reach comparable probabilities as PSF model 4. In particular, PSD model 2 (with a timescale of 154 min) shows a significantly improved likelihood with respect to model 1, but incorporates one additional parameter, the second break. 

Using PSD model 4, we
determined the acceptance value for the FDD by simulating 1000 23~h
stretches and fitting their CDFs in the same way as described in
section \ref{cdfsec}. By comparing the resulting $KS$ values to the
one derived from our observations, we found a $p$-value of $65\%$,
i.e., more than half of the realizations show higher $KS$-values than
the measured light curve.

\begin{figure}[ht]
   \centering
   \includegraphics[width=8.75cm]{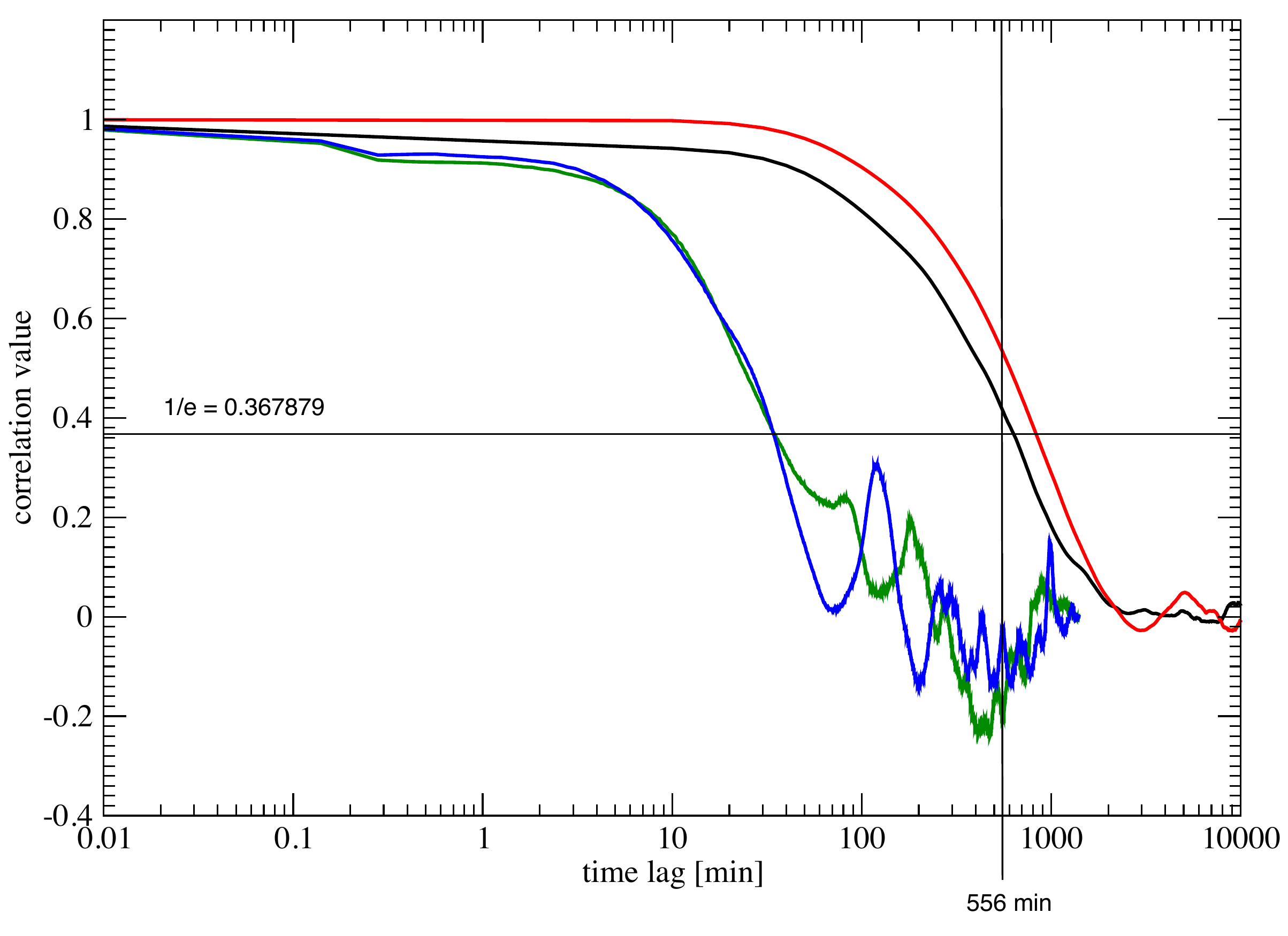}
   \caption{Autocorrelation functions of real and simulated data. The
    green line shows the autocorrelation function of the observed data,
    and the blue line shows that of the simulated light curve in the
    upper left panel of Figure~\ref{lc2}.  The red line shows the
    autocorrelation function corresponding to PSD model 4 derived from a 20000 minute Gaussian light curve with 0.1 minute sampling, and the black line is the analogous autocorrelation function derived from a power law distributed light curve. The vertical black line marks the 556 minute break, and the horizontal black line marks the 1/e level, which is sometimes used as an alternative characteristic value to describe the timing behavior (also referred to as the timescale).}
         \label{auto}
\end{figure}
\section{Discussion and Future Prospects}\label{s:discussion}

The statistical analysis shows that the data taken with \Sp\ at
4.5~$\mu$m are fully consistent with the $K$-band light curves. The offset
$c = 0.94$~mJy is a plausible value for restoring the average Sgr A* flux density that was lost by our reduction method. The scaling factor
$s = 1.0$ suggests that the observed flux density at 4.5~$\mu$m is
a good estimator of the dereddened flux density at 2.2~$\mu$m. The
derivation of $s$ was based on the observed fluctuation amplitudes and is
independent of prior knowledge of the source spectral index
$\alpha = 0.6$ derived from
synchronous measurements in the range of 1.6~$\mu$m to 3.7~$\mu$m and
the 4.5~$\mu$m extinction $A_{M} = 1.0 \pm 0.3$~mag (\citealt{2011A&A...532A..83S}). These
earlier observations predict  $s = 0.6^{+0.5}_{-0.3}$, and our value $s=1.0$ lies
within the bright end of the  $1\sigma$~range. This implies that the spectral properties
previously found, in particular a spectral index $\alpha=0.6$ characteristic of
optically thin synchrotron radiation, are valid for the
extended range of 1.6~$\mu$m to 4.5~$\mu$m and that the flux density
calibration worked consistently in both bands.

With the scaling factor known, we can characterize the sensitivity of
the \Sp\ measurements for the intrinsic variability of \Sg. The
typical noise level of observations with VLT/NACO is
$0.32$~mJy in the dereddened
light curves with a cadence of about 1.2 minutes  (\citealt{2012ApJS..203...18W}) .  The noise level  in the  8.4-s cadence 4.5~\micron\ BCD coadds before dereddening is about 0.7~mJy. If we block-average the BCD coadds  
over 7 points to create a cadence of about one minute,  the $S/N$ of the \Sp\ data is 0.25~mJy, a factor of $\sim$1.3 better than the  average ground-based AO observations with VLT/NACO.

The timing analysis shows that the observed 4.5~\micron\ light curve is
consistent with the existing model for the FDD and PSD as derived from $K$-band measurements. 
Based on the flux density distribution measured with the VLT, the 23~h \Sp\ light
curve helps discriminate between proposed models for
the PSD beyond what could be achieved by continued ground based
observations only: in the analysis of
\cite{2012ApJS..203...18W}, the likelihood ratio between a timescale of 154 minutes and PSD slope of 2.0 ($p(\chi^{2})=75\%$) and the model with the highest likelihood (break at 588 minutes, $p(\chi^{2})>94\%$) is 1.25. In our analysis, the  likelihood ratio between PSD 1 ($p(\chi^{2})=41\%$) and PSD 4 ($p(\chi^{2})=74\%$) is 1.8. 
The continuity of the sampling allowed us to develop a metric that is even more sensitive, the FDD-constrained likelihood $p(\chi^{2}\vert\rm{CDF}) $, in which PSD 1 is estimated to be 51 times less likely than PSD 4. Under these FDD constraints, the observed structure function maximum values between the time lags of 400 minutes and 600 minutes are beyond the $99.7\%$-confidence levels of structure functions generated with PSD 1.

The correlation timescale influences the length of the gaps between flux outbursts and the width of the flux peaks. A longer timescale means longer gaps and wider peaks, a shorter timescale means shorter gaps and narrower peaks. To further illustrate differences between various PSDs,  Figures~\ref{lc}
and~\ref{lc2} show example light
curves created with PSD models 1 and 4. The model with the
lower timescale shows many smaller outbursts but fails to
reproduce long stretches of flux densities very close to the
baseline.  The model
with the longer break timescale matches  the visual impression of
the observed data, showing rare large excursions clustered in time\footnote{A 560 minute break timescale corresponds to two or three periods of increased IR flux every 24 hours.}. Light curves
covering only $10-20\%$ of a day are not sensitive to this
difference. Figure~\ref{auto} shows the autocorrelation function
of the data in comparison to simulated 24 hour light curves, and  model 4
gives an excellent fit in this metric as well.

The analysis here is not yet able to rule out break timescales as short as 154 minutes due to the strong correlation between slopes and break timescales. PSD 2 (154 minute break combined with a steeper slope of 3.8 for timescales shorter than a second break at 20 minutes) is only 3.4 times less likely than PSD 4 and satisfies the  $99.7\%$-confidence levels for all time lags $\le$ 700 minutes. However, additional continuous $\sim$23~h monitoring will contribute essential information. 
Future observations of \Sg\ with \Sp/IRAC are planned for 2014
June--July (three epochs) and November--December (two epochs). These
will allow independent verification of the FDD model and a rigorous
determination of the break time scale and its uncertainty based on simultaneous Bayesian fitting of the CDF and structure function. They
should also show whether a second break (at 20 minutes) in the PSD is
warranted.
 
The variability of \Sg\ as a measure of the response of its accretion
flow to G2 is one of the key observables that can elucidate the
physics close to the event horizon. The latest orbital estimates put G2's time of closest
approach around 2014 March ($2014.21 \pm 0.13$;
\citealt{2013arXiv1312.1715M}; \citealt{2013ApJ...773L..13P};
\citealt{2013ApJ...774...44G}), which is just when the Galactic
center becomes observable again from ground-based telescopes (from
roughly October to February the Sun does not permit \Sg\ observations in the IR from
Earth). The light curve  reported here therefore fills a crucial gap;
these were the only IR observations possible in 2013 December, just before
G2's closest approach. Our analysis shows that G2 had not yet had a
measurable impact on \Sg: the statistical properties are exactly
as expected from more than a decade of $K$-band observations
(\citealt{2012ApJS..203...18W}, \citealt{2014arXiv1403.5289M}).
Our observations this year should show whether and how \Sg\
reacts to the presence of G2. 

\acknowledgments
This work is based on observations made with the \Sp\ Space
Telescope, which is operated by the Jet Propulsion Laboratory,
California Institute of Technology under a contract with
NASA. Support for this work was provided by NASA through an award
issued by JPL/Caltech. A.~G.\ acknowledges support from NSF grant  AST 09-09218. G.~W.\ acknowledges the European Union funded COST Action MP0905: Black Holes in a violent Universe and PECS project No. 98040.  We thank the staff of the \Sp\ Science Center for their help in 
planning and executing these demanding observations. We thank Keith Matthews and Arno Witzel for fruitful discussions.

Facilities: \facility{Spitzer/IRAC}

%\appendix
\begin{center}
 
APPENDIX A\\
\uppercase{
Observation Design and Data Reduction
}
\end{center}

 Each IER began
with a ``PCRS peakup'' offset from the star HD~316224 ($V = 10.2$; located
7\farcm0 from \Sg\ with accurate proper motions from {\it HIPPARCOS})
to place \Sg\ (R.~A.=$17^{\rm{h}}45^{\rm{m}}40^{\rm{s}}.036$, Decl.=$-29\arcdeg00^{\rm{m}}28^{\rm{s}}$.17,
  J2000; \citealt{2011AJ....142...35P}) on pixel (16,16) of the subarray (with the
coordinate of the first pixel in the subarray being 1,1).  Most
observations used a frame time of 0.1~s (thus 6.4~s duration for each
frame set), but some with a frame time of 0.02~s (1.28~s duration frame set)
were obtained as well (see below for details). The peakup offset
placed \Sg\ within 0.07 pixel of the desired position at the start of
each IER, but the pointing varied during the subsequent monitoring
observations, as described below.

The first IER (the ``mapping IER", AORKEY 50123264) made a small map
with maximum commanded offsets of 1\farcs1 in each coordinate of the
detector array. At each map position, observations were taken with
frame times of both 0.02 and 0.1~s. The shorter frames have no
saturated sources whereas the longer ones have three sources
saturated.  All three saturated sources are well away from \Sg\ and
do not affect the flux measurements, but we were concerned that they
might affect the determination of the pointing position of the
image.  This concern proved unfounded, and positions determined from
consecutive 1.28-s and 6.4-s frame sets agree to within 0.011 pixel rms.

The second and third IERs of the campaign (the ``monitoring IERs'',
AORKEYs 50123520 and 50123776) each consisted of an initial PCRS
peakup, one 1.28-s frame set, 5000 6.4-s frame sets, and a final
1.28-s frame set. Except for the initial PCRS peakup offset, no telescope motion was commanded during these IERs. Frame
sets were generally separated by 2~s of spacecraft overhead, resulting in an observation cadence of 8.4~s. 
The gap between the end of the 
second IER and the start of the IRAC data collection in the third IER was about 3.8~minutes (the gap was required for spacecraft overhead and for the second PCRS peakup operation). The entire monitoring
campaign began at JD~2456636.6802 (solar system barycenter) and ended
at JD~24456637.6551, a duration of 23.4~hours. The start time
corresponds to 2013 December 10 at 04:20:19~UTC.

\begin{figure*}
\begin{center}
\includegraphics[scale=0.60, angle=0]{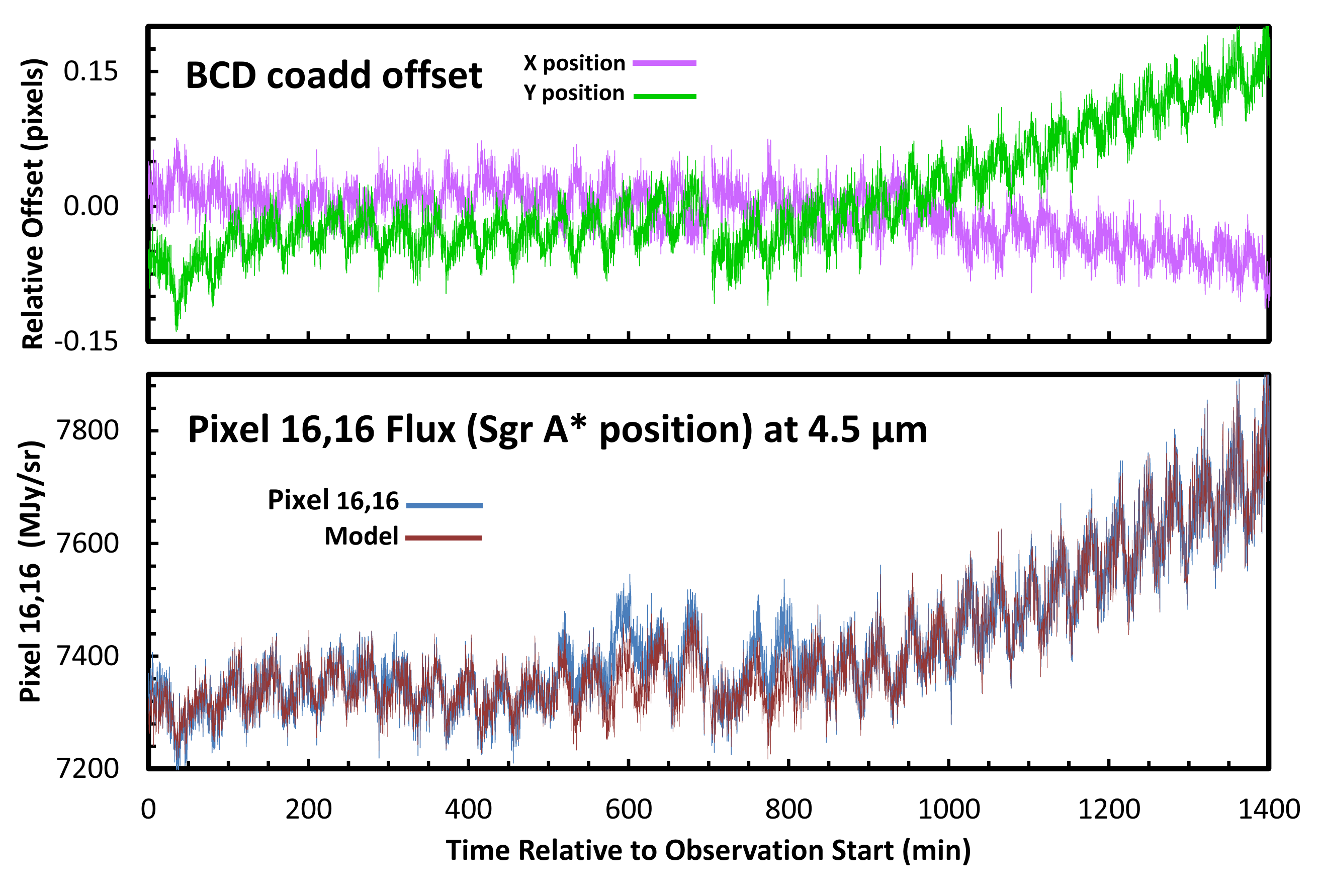}
\end{center}
\caption{ Top panel: $(X,Y)$ offsets in pixels for  the two
  monitoring observations.  The purple line shows
  the $X$ position and the green line the $Y$ position of \Sg\ relative to the center of pixel 
  (16,16).  The break near the center of the plot is where the
  peakup operation occurred between the first and second monitoring
  IERs.  Bottom panel: Value of pixel (16,16) in the 6.4-s BCD coadds, in units of MJy~sr$^{-1}$. The blue line shows
 the data, and the red line is the polynomial model fit described in the text (equation~\ref{polyfunc}). The
  horizontal axis shows the time in minutes relative to the start
  time of the first monitoring BCD. The panels have the same horizontal scale. }\label{pixoutpos}
\end{figure*}

 After constructing the 6.4-s BCD coadds as described in \S 2, the next reduction step was to derive their accurate relative positions.  We used the {\tt xregister} routine in
IRAF\footnote{IRAF is distributed by the National Optical Astronomy
  Observatory, which is operated by the Association of Universities
  for Research in Astronomy (AURA) under cooperative agreement with
  the National Science Foundation.} to perform a cross-correlation
between image pairs to determine the relative offset.  In order to
accurately register the subarray frames and correct the World
Coordinate System (WCS) defined in the FITS headers of each
BCD\footnote{The small field of view of the subarray includes too few
  2MASS reference stars to permit the standard pipeline processing to
  derive a reliable WCS for our IERs, and therefore the RA and Dec
  positions in the Galactic center frame set headers have
  significantly larger uncertainties than typical full-frame IRAC
  BCDs.}, we first performed a cross-correlation between the first
0.1~s BCD coadd and each of the mapping IER BCD coadds to determine the
relative offsets, and wrote the relative RA and Dec into the BCD coadd
headers. We then used IRACproc to construct a mosaic using these
frames at a resolution of 0\farcs6 pixel$^{-1}$.  The RA and Dec of
this mosaic were then determined by comparison to a ground-based {\it
  L}-band adaptive optics image obtained with NIRC2 at Keck (see
Figure \ref{f:map}). The 4.5 $\mu$m IRAC mosaic was then re-projected back to the
instrumental pixel scale with \Sg\ placed at the center of pixel
(16,16).  The reprojected mosaic was then used as a reference image,
and the relative offset of each of the monitoring BCD coadds was
determined by cross-correlation with this reference image. The
derived relative offsets for the monitoring BCD coadds are shown in
the top panel of Figure \ref{pixoutpos}. The \Sp\ spacecraft
battery-heater-related pointing oscillation (\citealt{2012SPIE.8448E..1IG}) is
visible throughout the observation. It has a period of $\sim$40
minutes and a peak-to-peak amplitude of $\sim$0.05 pixels in X and Y.
The mean pointing was relatively constant during the first monitoring
IER, but during the second there was a roughly constant drift
throughout the observation, resulting in an offset of $-0.10$ pixel in
X and +0.25 pixels in Y by the end of the 700 minute IER.

Because the intrinsic IRAC response is extremely stable, the pixel output depends 
only on the position of the frame on the sky
and the variability of \Sg. This is similar but not identical to
the case of exoplanet transit  observations, where pointing
variations cause a single point source to move relative to the pixel
center (e.g., \citealt{2010PASP..122.1341B}; \citealt{2014arXiv1405.3659B}). In the exoplanet case,  a single
bright point source dominates and the pointing variations cause changes in the measured signal as result of the structure of the IRAC intra-pixel response, which requires a multiplicative correction. In the
  Galactic center case, the major correction is additive because the pointing-induced signal variations are due to the IRAC pixel sampling different parts of the complex background of unresolved sources and extended emission near \Sg.

The bottom panel in Figure \ref{pixoutpos} shows the effect of the
pointing variations on the pixel output of the coadded BCD frame
sets.  In addition to the position-dependent variations, the output
includes the noise inherent in the observation and the variability of
\Sg\ itself. The median surface brightness is 7385 MJy sr$^{-1}$ (or
257 mJy pixel$^{-1}$), and the maximum surface brightness gradient is
1600 MJy sr$^{-1}$ arcsec$^{-1}$ (56 mJy pixel$^{-1}$  arcsec$^{-1}$
).

We experimented with several methods for extracting the \Sg\ flux density, including simple aperture photometry, difference imaging plus aperture photometry, and using various linear combinations of the eight pixels adjacent to (16,16) to adjust the flux density.  No method gave lower noise during the low parts of the light curve than simply using pixel (16,16) alone, corrected as described below. Some of the other methods resulted in large artifacts (many times the level of the signal) that were strongly correlated to the frame position, clearly failing to remove the effects of the pointing variations. From standard star
measurements we know that $\sim$40\% of the flux from a
well-centered point source will fall within the central pixel, and
less than 10\% will fall in the next brightest neighboring pixel.  The pixels
near \Sg\ sample the complex field of bright, unresolved point
sources and significant extended emission in the Galactic center
region, so they all vary strongly as a function of
position. Therefore, adding neighboring pixels to the analysis
contributes more systematic errors than additional signal and does
not improve the signal to noise of the measurement.  

To model the dependence of the pixel output on the X,Y position of \Sg\ in the frame, we used a
second-degree polynomial
\begin{equation}
F(X,Y) = a + bX + cY + dXY + eX^2 + fY^2
\label{polyfunc}
\end{equation}
and determined the coefficients by performing a least-squares fit to
minimize the residuals between the function and the monitoring data.  The first
iteration showed that there were time intervals where the data deviated significantly from
the fit; these regions were then excluded from the fit and the
coefficients determined again.  The final fitted values of the
coefficients are $a=7379.8$, $b=1196.8$, $c=-1096.5$, $d=-197.2$,
$e=-1948.4$, and $f=4704.6$.  The model value is plotted at each of the positions
observed along with the pixel (16,16) output in the bottom
panel of Figure \ref{pixoutpos}. 

The initial measure of the variable component of the flux density
from \Sg\ is the residual between the pixel (16,16) data value and
the expected value determined by equation~(\ref{polyfunc}).  These values
have to be multiplied by the position-dependent ratio
of total flux density to central pixel signal for a point source in order to determine the
total variable component of the flux from \Sg. The
 flux from a point source is distributed across IRAC pixels
according to the pixel response function (PRF). The  IRAC calibration was determined using
aperture photometry of stars (\citealt{2005PASP..117..978R}), so we used existing
observations of a standard star (BD+$67\degr$~1044) taken in subarray
mode with 0.1 s frame times to determine the relationship between central
pixel flux and centroid position of a point source, relative to the total source flux. 
We used $\sim$18,800 warm mission subarray
measurements originally taken to map out the subarray response for
exoplanet observations. Standard star data with centroids ranging
from $-0.1$ to $+$0.2 in $X$ and $-0.2$ to $+$0.2 in $Y$ (relative to
the center of pixel 16,16) were used to cover the range of pixel
coordinates seen in the \Sg\ data. A fourth-degree polynomial as a
function of $X,Y$ position was needed to fit the central pixel to
total flux ratio. This polynomial reproduces the measurements to an
accuracy of 0.04\%~rms. The central value of the fitted function is
0.407, and other values range from 0.335 to 0.430 over the
observed range of \Sg\ positions. The final calibrated light curve data are given in Table~\ref{LightcurveData},
and plotted in Figure~\ref{sgrAplot}.

As a test of our reduction method, we also extracted and modeled the
output of pixel (18,19) in the same way as for (16,16). This pixel is
on an image location with a significant gradient and not on a local
maximum, similar to pixel (16,16) but far enough away from it that
it will not see the variability from \Sg\ (see Figure~\ref{f:map}). The
median value in this pixel is ${\sim}10^4$~MJy sr$^{-1}$ (350~mJy
pixel$^{-1}$). The fit coefficients as in eq.~\ref{polyfunc} for pixel (18,19) are
  $a=10064.6$, $b=6686.8$, $c=-4260.6$, $d=5658.8$, $e=-4263.9$, and
  $f=-761.7$.  The results are plotted in Figure \ref{sgrAplot}.
There is one peak near the beginning of the time series that rises to
$\sim$2.5 mJy in the smoothed data, but other fluctuations are
smaller, and the rms is similar to that for pixel (16,16). The reference pixel data
are also given in Table~\ref{LightcurveData},
and plotted in Figure~\ref{sgrAplot}.

\begin{center}
APPENDIX B\\
\uppercase{
Statistical Analysis Methods for the Power Spectral Density
}
\end{center}

Predicted structure functions offer a way to test PSD models.  For each
proposed PSD, we derived a set of structure functions from
simulated light curves. We used the algorithm from
\cite{2012ApJS..203...18W} based on the method by
\cite{1995A&A...300..707T} to create light curves that exhibit the power-law FDD
described in Section~\ref{cdfsec}. The procedure began by drawing
Fourier coefficients for each frequency from a Gaussian distribution
with a variance proportional to the value of the PSD at that
frequency. The resulting PSD was Fourier-transformed to the time
domain and normalized to unity variance. At this stage, the equally
spaced data were optionally resampled to the cadence of the observed light
curve.  With or without resampling, each point was transformed according to
equation~(\ref{transf}).  Finally, an independent Gaussian
noise was added to each point to account for measurement errors.  For
all models tested here, the best result for the first time lag was
achieved by giving the white noise a standard deviation of 0.7~mJy.

\begin{figure}[ht]
   \centering
   \includegraphics[width=8.6cm]{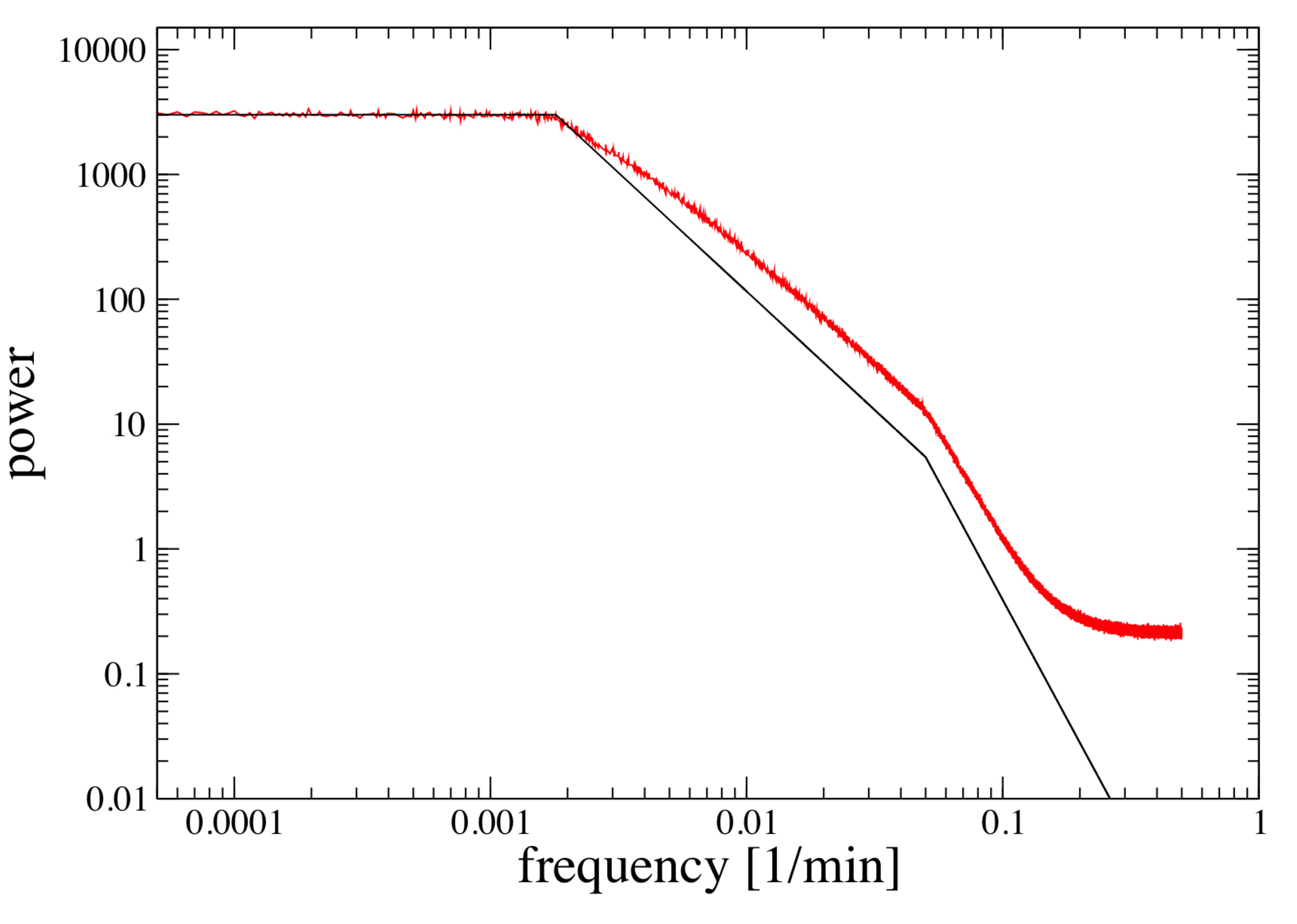}
   \caption[PSD model.]{PSD versus temporal frequency for model 4 (Table \ref{PSDparams}). The black line shows the input PSD that creates the Gaussian time series. The red line shows the PSD derived from 10000 simulated light curves with a length of 20000 minutes each, sampled at 0.1 minute intervals. The difference between the two lines arises because the initial light curve created from the PSD is Gaussian, but the source fluctuations are modeled as a power law distribution. The flattening towards highest frequencies is caused by the white measurement noise. The break locations are the same for both distributions, and are given in Table \ref{PSDparams}.}
         \label{psd}
\end{figure}

The important property of the algorithm for generating predicted structure
functions is the normalization step. It ensures that for any break
timescale and PSD slope, the PSD is normalized in a way that flux
densities occur with the observed probabilities. In particular, it
enables us to compare the absolute values of the structure function
of simulated light curves with the observed structure function. The
transformation changes the PSD of the generated light curve slightly
(see difference between input PSD and effective PSD in Figure~\ref{psd}
illustrated for a double broken PSD). The break timescales, however,
are not affected. 

\begin{figure}[ht]
   \centering
   \includegraphics[width=8.6cm]{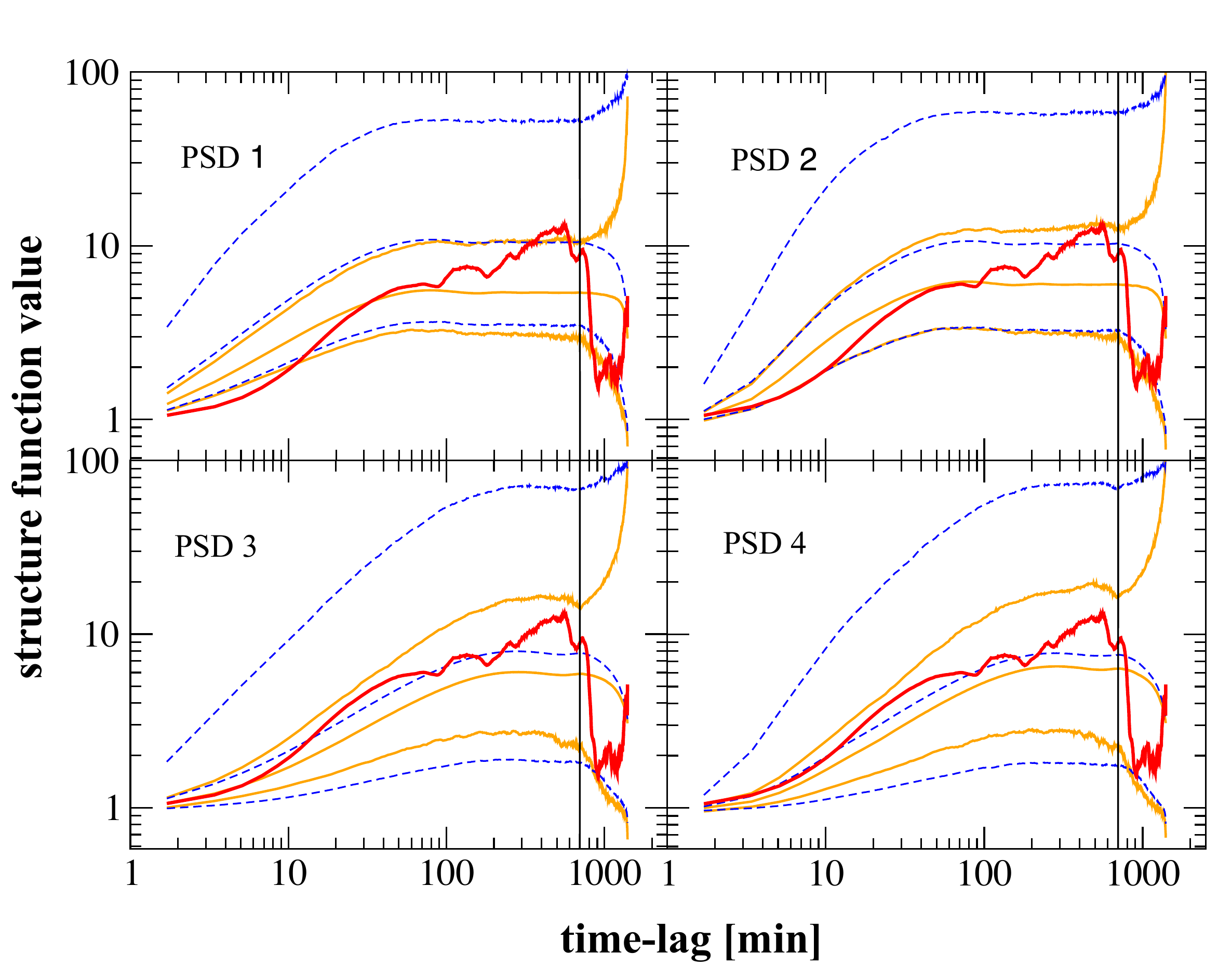}
   \caption[Confidence intervals for structure function.]{
     The structure function and confidence intervals for our four PSD models. The thick red lines show the observed data. The dashed blue lines show the most probable structure function values and their 95\% confidence intervals for light curves without constraints. The orange solid lines represent the most probable structure function values and their 99.7\% confidence intervals for light curves with matching CDFs (for all models the CDFs differ from the observed data by $< 0.07$ and the peak flux density over 23.4 h differing by $< 2.5$ mJy).  The vertical lines mark the time lag corresponding to half the monitoring duration.
     }         \label{conf}
\end{figure}

For each of the PSD models, we generated 10000 light curves with a sampling of
0.1~minutes and a length of 20000~minutes (see
Figure~\ref{conf}), resampled a 23~h middle section
(avoiding the ends in order to avoid red-noise leakage) to the actual
cadence of the light curve, and
calculated structure functions and the most probable structure
function value for each time lag $\tau_m$: 
\begin{equation}
\tilde{V}(\tau_{m}) = \exp{\langle\ln[V(\tau_{m})]\rangle}~~,
\end{equation}
and its $95\%$- and $99.7\%$-confidence levels (by determining the 0.135\ts{th}, 2.275\ts{th}, 97.725\ts{th}, and 99.865\ts{th} percentile).  We also calculated the pseudo-$ \chi^{2} $
(\citealt{2010MNRAS.404..931E}): 
\begin{equation}
\chi^{2}_{\rm{ps}} =
\sum_{m}\left(\frac{\langle\ln[V(\tau_{m})]\rangle -
    \ln[V(\tau_{m})]}{\sigma_{m}} \right)^{2}~~, 
\end{equation}
with $ \sigma_{m} $ the  variance of $ \ln[V(\tau_{m})]$.
Pseudo-$ \chi^{2}$ quantifies the deviation from the most probable values,
$\tilde{V}(\tau_{m})$, summed over all time lags and defines a likelihood function for the PSD parameter set by determining the fraction of simulated structure functions with a worse $\chi^{2}_{\rm{ps}}$ than the observed data. The results for
the four PSD models are shown in Figure~\ref{conf}. 

For every tested PSD model, the $95\%$-confidence level is wide
enough to account for the deviations of the observed structure
function from the most probable values up to a time lag of 700 minutes. The longer-timescale
PSD models have likelihood values of $p(\chi^{2}_{\rm{ps}})=73\%$ and $p(\chi^{2}_{\rm{ps}})=74\%$, whereas the 154~minute PSDs have likelihood values of $p(\chi^{2}_{\rm{ps}})=41\%$ and $p(\chi^{2}_{\rm{ps}})=44\%$. However, testing the  structure functions against the 
general set of 23~h stretches generated from a particular PSD does not take
advantage of all the information. The amplitude of typical flux density fluctuations scales with the flux density level itself (rms-flux relation). Thus, letting light curves  that exhibit very different FDDs over 23~h 
contribute to the statistics of the structure function is
not acknowledging that the measured FDD has a specific flux density
maximum.

To take full advantage of the data, we defined a modified likelihood function based on a restricted set of light curves that have a maximum similar to that seen in the data (10.3$\pm
2.5$~mJy) and a similar CDF (maximum difference between observed
and simulated CDF $<$0.07). This likelihood function was defined as the product of the probability $p(\rm{CDF})$ to find the observed FDD and the probability to find a structure function with a larger $\chi^{2}_{\rm{ps}}$ value, $p(\chi^{2}_{\rm{ps}}\vert\rm{CDF}) $.  

Additionally, the constraints on the FDD allow us to determine whether the measured structure function values are in the range of the statistical expectation for the observed flux densities. Introducing flux density constraints makes sense only for continuous datasets without gaps larger than the time lag binning. Otherwise, depending on the observation gaps, the FDD from which the individual time lag draws changes, and the constraints do not affect each time lag equally. The $\chi^{2}_{\rm{ps}}$ value in this case has to be computed from time lags $\le 700$~minutes only (which draw from the full number
of data points in the light curve), and the confidence
intervals derived in this way are strictly correct only for
those shorter time lags (the FDD of flux densities
contributing to structure function values at higher time lags would
have to be separately matched with the simulation, time lag by
time lag, requiring excessive computation time).
\bibliographystyle{apj}
\bibliography{mybib_gc}{}

\end{document}